\documentclass[a4paper,11pt]{article}

\usepackage{pos}
\usepackage{tikz}
\usepackage{amsmath}
\usepackage{amsfonts}
\usepackage{amsthm}
\usepackage{lineno}

\newcommand{\mr}{\mathrm}
\def\MS{\overline{\rm MS}}
\newcommand{\al}[1]{\vskip-3ex\begin{align}#1\end{align}}

\title{Update on (2+1+1)-flavor QCD equation of state}

\author*[a]{Johannes Heinrich Weber}
\author[b]{Alexei Bazavov}
\author[c]{Peter Petreczky}

\affiliation[a]{Institut f\"ur Physik \& IRIS Adlershof; Humboldt-Universit\"at zu Berlin,\\
Zum Gro\ss en Windkanal 6, D-12489 Berlin, Germany}

\affiliation[b]{Department of Computational Mathematics, Science and Engineering \& Department of Physics and Astronomy; Michigan State University,\\
428 S. Shaw Ln., East Lansing, MI 48824, USA}

\affiliation[c]{Physics Department; Brookhaven National Laboratory,\\
Bldg. 510, Upton, NY 11973-5000, USA}

\emailAdd{johannes.physik@hu-berlin.de}

\abstract{We report on preliminary results from the calculations of the QCD equation of state
for 2+1+1 flavors using HISQ action. The calculations are performed on lattices with temporal
extents $N_{\tau}=6,~8,~10$, and $12$ and aspect ratio $N_{\sigma}/N_{\tau}=4$. We find that there is a significant
contribution to the pressure from charm quarks at temperatures $T> 300$ MeV.}

\FullConference{%
 The 38th International Symposium on Lattice Field Theory, LATTICE2021
  26th-30th July, 2021
  Zoom/Gather@Massachusetts Institute of Technology
}

\begin{document}
\maketitle

\section{Introduction}

The quark-gluon plasma (QGP), the high-temperature phase of bulk nuclear matter, 
has been studied in ultra-relativistic heavy-ion collision (HIC) experiments at 
RHIC (BNL), LHC (CERN) for many years, and will be probed after their upgrades 
and in future experiments such as FAIR (GSI) and NICA (JINR), too. 
At vanishing baryon density the transition between the hadron gas and the QGP 
takes place as a broad chiral crossover around a temperature of 
$T_\mr{pc}=156.5(1.5)\,\mr{MeV}$ at the physical point~\cite{HotQCD:2018pds}.
The thermodynamic properties of QGP are given in terms of its equation of state 
(EoS), which has been studied extensively on the lattice 
in pure gauge theory (without sea quarks)~\cite{Giusti:2016iqr}, 
or with 2+1 dynamical flavors (\textit{i.e.} light quarks in the isopin limit, and a physical 
strange quark) of sea quarks~\cite{Borsanyi:2013bia, HotQCD:2014kol, Bazavov:2017dsy}; 
after clearing up discrepancies between early lattice calculations 
due to a poorly controlled continuum limit, 
good agreement was achieved in (2+1)-flavor QCD. 

Heavy quarks are negligible in nuclei. 
Instead, they are produced in hard processes during early stages of the HIC. 
Future HIC experiments at larger $\sqrt{s}$ will lead to higher temperature 
and copious production of charm. 
Furthermore, for physics of the early universe the charm contribution to the
equation of state cannot be neglected, see e.g. Ref. \cite{Borsanyi:2016ksw}.
Thus it is urgent to include dynamical charm quarks in the lattice calculation
of the equation of state.
Heavy quarks are challenging due to the large discretization errors associated 
with their mass, see e.g. the difficulty of the continuum limit for moments of 
pseudoscalar charmonium correlators~\cite{Petreczky:2020tky}. 
At $T \gtrsim 2T_\mr{pc}$ the previously dominant gluon contribution 
and the light or strange quark contributions die down rapidly, 
whereas the contribution from charm quarks catches up as 
thermal scales, i.e. $\pi T$, approach its mass ($\MS$: 
$m_c(m_c,N_f=4)=1.2735(35)\,\mr{GeV}$~\cite{Komijani:2020kst}). 
Charm quarks give an important contribution to the EoS at temperatures for which 
weak-coupling calculations are not yet reliable \cite{Laine:2006cp}.
Although results in (2+1+1)-flavor QCD (\textit{i.e.} with a charm sea) have been 
obtained already some time ago~\cite{Borsanyi:2016ksw}, no independent cross-check 
through a calculation using another discretization for the charm sea is 
available yet. 
In this contribution we report on an ongoing (2+1+1)-flavor QCD study~\cite{MILC:2012aiy, MILC:2013ops}
with highly improved staggered quark (HISQ) action~\cite{Follana:2006rc} 
optimized for controlling heavy-quark mass discretization effects. 

\section{Lattice setup}

Any lattice calculation of the EoS is computationally demanding.
In the traditional approach that we follow, i.e. the integral method, 
both $T>0$ and $T=0$ ensembles with high statistics are needed at each 
bare gauge coupling to cancel UV divergences. 
We use coarse $T>0$ lattices with aspect ratio $N_\sigma/N_\tau=4$ and 
temporal extents $N_\tau=6,~8,~10$, and $12$;
the temperature is set as $T=1/(aN_\tau)$. 
The data set is anchored to a set of existing, high statistics MILC 
ensembles~\cite{Bazavov:2017lyh} at $T=0$ along the line of constant physics 
(LCP) with a light quark mass $m_l=m_s/5$, 
i.e. $m_\pi \approx 300\,\mr{MeV}$ in the continuum limit. 
We combine the HISQ action~\cite{Follana:2006rc} 
with a tadpole one-loop improved gauge action.  
HISQ suppresses taste exchanges and diminishes mass splittings in the pion sector;
this improves the approach to the continuum limit at low temperatures. 
HISQ is $O(a^2)$-improved at tree-level due the Naik (three-link) term, which
improves scaling at high temperatures~\cite{Bazavov:2017dsy}, 
and contains a mass-dependent correction $\epsilon_N$ for the charm quark~\cite{Follana:2006rc}, 
which reproduces the correct charm dispersion relation at tree-level up to $O((am_c)^4)$. 

We use the $r_1$ scale defined in terms of static potential at $T=0$ to set the lattice spacing $a$.
We use the value $r_1\simeq 0.3106\,\mr{fm}$ \cite{MILC:2010hzw} in this study.
Strange and charm quark masses are tuned to physical values by using masses of 
$\pi$, $K$, and the spin average of $\eta_c$ and $J/\psi$. 
The tadpole factor defined from the trace of the
plaquette $u_0=\left\langle{\rm Tr}~U_p/3\right\rangle^{1/4}$ is determined during
thermalization of the $T=0$ ensembles. 
The parameters and accumulated statistics for the $T=0$
ensembles are shown in Table~\ref{tab:statT0}.
Corresponding temperatures and the statistics for the $T>0$
ensembles are shown in Table~\ref{tab:statTne0}. 
We cover a window of $T \in [149,967]\,\mr{MeV}$ with $N_\tau=6$ 
and $T \in [136,725]\,\mr{MeV}$ with $N_\tau=8$. 
\begin{table}
\centering
\begin{tabular}{|l|l|l|l|l|l|r|}
\hline
$\beta$ & $V$ & $am_l$ & $am_s$ & $am_c$ & $a$, fm & TU \\
\hline
5.400  & $16^3\times40$  & 0.0182  & 0.091  & 1.339 & 0.220 & 20K \\
5.469  & $24^3\times32$  & 0.01856 & 0.0928 & 1.263 & 0.206 & 19K \\
5.541  & $24^3\times32$  & 0.01718 & 0.859  & 1.157 & 0.192 & 18K \\
5.600  & $16^3\times48$  & 0.0157  & 0.0785 & 1.08  & 0.181 & 69K \\
5.663  & $24^3\times32$  & 0.01506 & 0.0753 & 0.996 & 0.170 & 28K \\
5.732  & $32^4$          & 0.01394 & 0.0697 & 0.913 & 0.159 &  10K \\
5.800  & $16^3\times48$  & 0.013   & 0.065  & 0.838 & 0.151 & 99K \\
5.855  & $32^4$          & 0.01216 & 0.0608 & 0.782 & 0.140 & 15K \\
5.925  & $32^4$          & 0.01122 & 0.0561 & 0.716 & 0.130 & 14K \\
6.000  & $24^3\times64$  & 0.0102  & 0.0509 & 0.635 & 0.121 & 11K \\
6.060  & $32^4$          & 0.00962 & 0.0481 & 0.603 & 0.113 & 38K \\
6.122  & $32^4$          & 0.00896 & 0.0448 & 0.558 & 0.106 & 38K \\
6.180  & $32^4$          & 0.0084  & 0.042  & 0.518 & 0.100 &  38K \\
6.238  & $32^4$          & 0.00784 & 0.0392 & 0.482 & 0.095 &  40K \\
6.300  & $32^3\times96$  & 0.0074  & 0.037  & 0.44  & 0.089 &  6K \\
6.358  & $32^4$          & 0.00682 & 0.0341 & 0.416 & 0.089 &  9K \\
6.445  & $32^4$          & 0.00616 & 0.0308 & 0.374 & 0.077 &  15K \\
6.530  & $36^3\times48$  & 0.0056  & 0.028  & 0.338 & 0.070 &  11K \\
6.632  & $48^4$          & 0.00498 & 0.0249 & 0.300 & 0.063 &  3K \\
6.720  & $48^3\times144$ & 0.0048  & 0.024  & 0.286 & 0.058 &  6K \\
6.875  & $48^3\times64$  & 0.0038  & 0.019  & 0.228 & 0.050 &  3K \\
7.000  & $64^3\times192$ & 0.00316 & 0.0158 & 0.188 & 0.045 &  6K \\
7.140  & $64^3\times72$  & 0.0029  & 0.0145 & 0.172 & 0.039 &  4K \\
7.285  & $64^3\times96$  & 0.00248 & 0.0124 & 0.148 & 0.034 &  4K \\
\hline
\end{tabular}
\caption{Parameters of the calculations at zero temperature,
including, the lattice gauge coupling $\beta=10/g_0^2$, quark masses, lattice spacings as well
as the corresponding statistics in terms 
of molecular dynamics time units (TUs).}
\label{tab:statT0}
\end{table}

\begin{table}
\centering
\begin{tabular}{|l|l|r|l|r|l|r|l|r|l|r|}
\hline
$\beta$ & \multicolumn{2}{|c|}{$N_\tau=6$} &
\multicolumn{2}{|c|}{$N_\tau=8$} &
\multicolumn{2}{|c|}{$N_\tau=10$} &
\multicolumn{2}{|c|}{$N_\tau=12$}  \\
\hline
 & $T$ & TU & $T$ & TU & $T$ & TU & $T$ & TU \\
\hline
5.400 & 149 & 50K &     &     &     &     & & \\
5.469 & 160 & 50K &     &     &     &     & & \\
5.541 & 171 & 50K &     &     &     &     & & \\
5.600 & 182 & 50K & 136 & 114K &     &     & & \\
5.663 & 193 & 50K & 145 & 74K &     &     & & \\
5.732 & 207 & 50K & 155 & 86K &     &     & & \\
5.800 & 218 & 50K & 163 & 81K & 131 & 40K & & \\
5.855 & 235 & 50K & 176 & 105K & 140 & 42K & & \\
5.925 & 253 & 50K & 190 & 105K & 152 & 42K & & \\
6.000 & 272 & 50K & 204 & 105K & 163 & 40K & 136 & 39K \\
6.060 & 291 & 50K & 218 & 99K & 175 & 42K & 145 &  21K \\
6.122 & 310 & 50K & 233 & 101K & 186 & 42K & 155 & 21K \\
6.180 & 329 & 50K & 247 & 99K & 197 & 40K & 165 & 32K \\
6.238 & 346 & 50K & 260 & 96K & 208 & 13K & 173 &  27K \\
6.300 & 369 & 50K & 277 & 98K & 222 & 84K & 184 & 28K \\
6.358 & 391 & 50K & 294 & 96K & 235 &     & 196 & 4K\\
6.445 & 427 & 50K & 320 & 96K & 256 &     & 214 & 4K\\
6.530 & 470 & 50K & 352 & 99K & 282 &  59K & 235 & 10K \\
6.632 & 522 & 50K & 391 & 96K & 313 &     & 261 & \\
6.720 & 567 & 50K & 425 & 100K & 340 & 10K & 284 & 10K \\
6.875 & 658 & 50K & 493 & 108K & 395 &     & 329 & 11K \\
7.000 & 731 & 40K & 548 &  110K & 438 & 20K & 366 & \\
7.140 & 843 & 40K & 632 & 11K & 506 & 19K & 422 &  2K \\
7.285 & 967 & 40K & 725 & 11K & 580 & 17K & 483 &  2K \\
\hline
\end{tabular}
\caption{Statistics of $T>0$ calculations for different $N_{\tau}$ in terms of
molecular dynamics time units (TUs). Under each $N_\tau$ the first of the
two columns shows the temperature $T$ in MeV and the second the number of TUs.
}
\label{tab:statTne0}
\end{table}

\section{Trace anomaly}

In the standard approach the EoS is obtained from the trace 
of the energy-momentum tensor (EMT), 
$\Theta^{\mu\mu} = \varepsilon-3p$, where $\varepsilon$ or $p$ 
are energy density or pressure \cite{Boyd:1996bx}.
$\Theta^{\mu\mu}$ is related to 
the partition function as
\al{
\frac{\Theta^{\mu\mu}}{T^4} = -\frac{T}{V}\frac{d\ln Z}{d\ln a},\quad 
Z=\int DU D\bar\psi D\psi~ e^{-S_{g}-S_{f}}.
\label{eq:partfun}
}
The temperature-independent divergences of any individual contribution 
$X$ to $\Theta^{\mu\mu}$ can be 
removed by subtracting the vacuum result for this operator $X$, i.e. 
\al{
  \Delta(X)=\langle X \rangle_\tau - \langle X \rangle_0\,.
\label{eq:subtr}
}
The vacuum-subtracted trace anomaly is given in terms of the basic ingredients of the action,
\al{
  \frac{\Theta^{\mu\mu}}{T^4}
  &=-R_\beta(\beta)\left[\Delta(S_{g})+
  R_u(\beta)\Delta\left(\frac{dS_g}{du_0}\right)\right]
  + R_\beta(\beta) R_{m_s}(\beta)
  \left[2m_l\Delta(\bar\psi_l\psi_l)+m_s\Delta(\bar\psi_s\psi_s)\right]\nonumber\\
  &+R_\beta(\beta) R_{m_c}(\beta)
  \left[m_c\Delta(\bar\psi_c\psi_c)+
  R_{\varepsilon_N}(\beta)\Delta\left(\bar\psi_c
  \left[\frac{dM_c}{d\varepsilon_N}\right]\psi_c\right)\right]\,,\label{eq:e3p_all}
}
after the lattice spacing derivatives have been rephrased in terms of $\beta$ functions 
and action parameter derivatives. 
Changes of the lattice spacing and the action parameters along the LCP
are controlled by lattice $\beta$-functions:
\al{
R_\beta(\beta) &= T \frac{{\rm d} \beta}{{\rm d}T} =
 - a \frac{{\rm d} \beta}{{\rm d}a} = (r_1/a)(\beta) \,
 \left( \frac{{\rm d} (r_1/a)(\beta)}{{\rm d}\beta} \right)^{-1} \, ,
\label{eq:Rb}\\
R_{m_q}(\beta) &= \frac{1}{am_q(\beta)} \frac{{\rm d} am_q(\beta)}{{\rm d}\beta}
\qquad \text{for} ~ q=s, c \,
\label{eq:Rm}, \\
R_u(\beta) &= \beta\, \frac{{\rm d} u_0(\beta)}{{\rm d}\beta}
\, , 
\qquad
R_{\epsilon}(\beta) = \frac{{\rm d} \epsilon_N(\beta)}{{\rm d}\beta} \, .
}
We have determined the $\beta$-functions by fitting the data to 
the following Allton-type Ans\"{a}tze~\cite{Allton:1996dn}.
For the lattice spacing:
\al{
\frac{r_1}{a}(\beta) = \frac{c_r^{(0)} f(\beta) + c_r^{(2)} (10/\beta) f^3(\beta)}
  {1 + d_r^{(2)} (10/\beta) f^2(\beta)} \, ,
\label{eq:ar1_para}
}
and for the strange or charm quark masses ($q=s,c$):
\al{\hskip-1ex
am_q(\beta) = \frac{c_q^{(0)} f(\beta) + c_q^{(2)} (10/\beta) f^3(\beta)}
  {1 + d_q^{(2)} (10/\beta) f^2(\beta)} \,
\left(\frac{20b_0}{\beta}\right)^{\frac{4}{9}} \hskip-1ex.
\label{eq:am_para}
}
Here $f(\beta)$ is the universal two-loop $\beta$-function for $N_f$ massless flavors
\al{
f(\beta) = \left(\frac{10b_0}{\beta}\right)^{-b_1/(2b^2_0)}
 \exp{(-\beta/20b_0)}.
\label{eq:beta_fn}
}
The obvious problem is that the charm quark mass can neither be neglected nor assumed
to be very large compared to the typical QCD scale. Therefore, we can only set $N_f=3$ or
$N_f=4$ and check for possible differences in the resulting parameterization of $r_1$ 
and the running quark masses. We used $N_f=3$ in the final result but
checked that using the $N_f=4$ the parameterization 
in Eqs.~\eqref{eq:ar1_para} 
and~\eqref{eq:am_para} would give statistically consistent results (although with different parameters).
To obtain the $\beta$ derivatives in Eq. \eqref{eq:Rm}, we 
fit $u_0$ with $u_0(\beta)=c_1+c_2e^{-d_1\beta}$ 
and $\epsilon_N$ with a polynomial in $\beta$. 

To obtain the pressure we use thermodynamic identity and write
\begin{equation}
\label{eq_pT4}
\frac{p(T)}{T^4} = \frac{p_0}{T_0^4} + \int_{T_0}^T dT'\frac{\Theta^{\mu\mu}}{T^{\prime5}},
\end{equation}
where $p_0$ is the pressure at some low reference temperature $T_0$. 
This is the integral method for calculating the pressure \cite{Boyd:1996bx}.
If we choose $T_0$ well
below the crossover temperature we can use the hadron resonance gas (HRG) model to evaluate $p_0$.
In our calculation we use the HRG model corresponding to the pion mass of $300$ MeV, which also
takes into account the taste splitting in the pion sector \cite{Bazavov:2017dsy}.

\section{Numerical results}

\begin{figure}
\includegraphics[width=14cm]{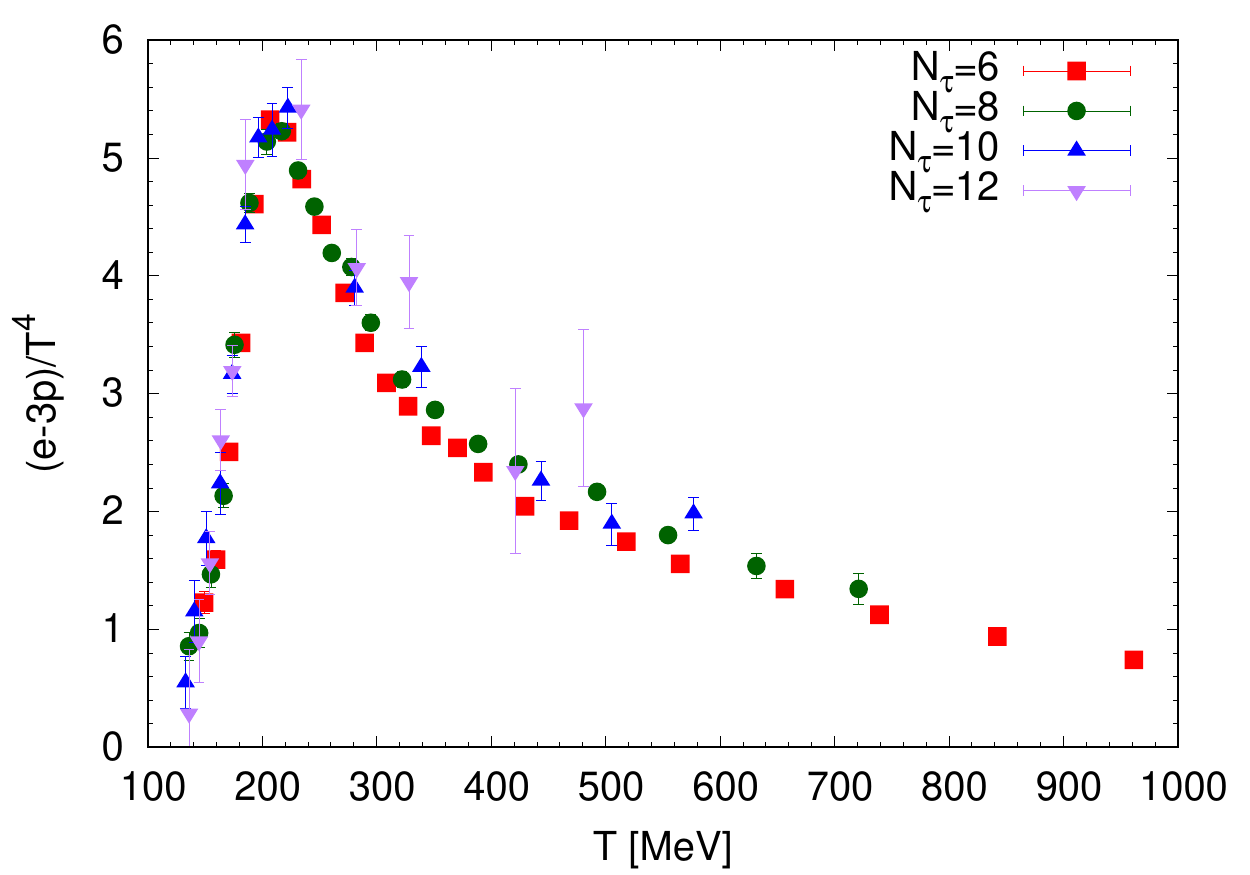}
\caption{The trace anomaly as function of the temperature calculated on $N_{\tau}=6,~8,~10$ and $12$ lattices.}
\label{fig:e-3p}
\end{figure}

\begin{figure}\centering
\includegraphics[scale=0.5]{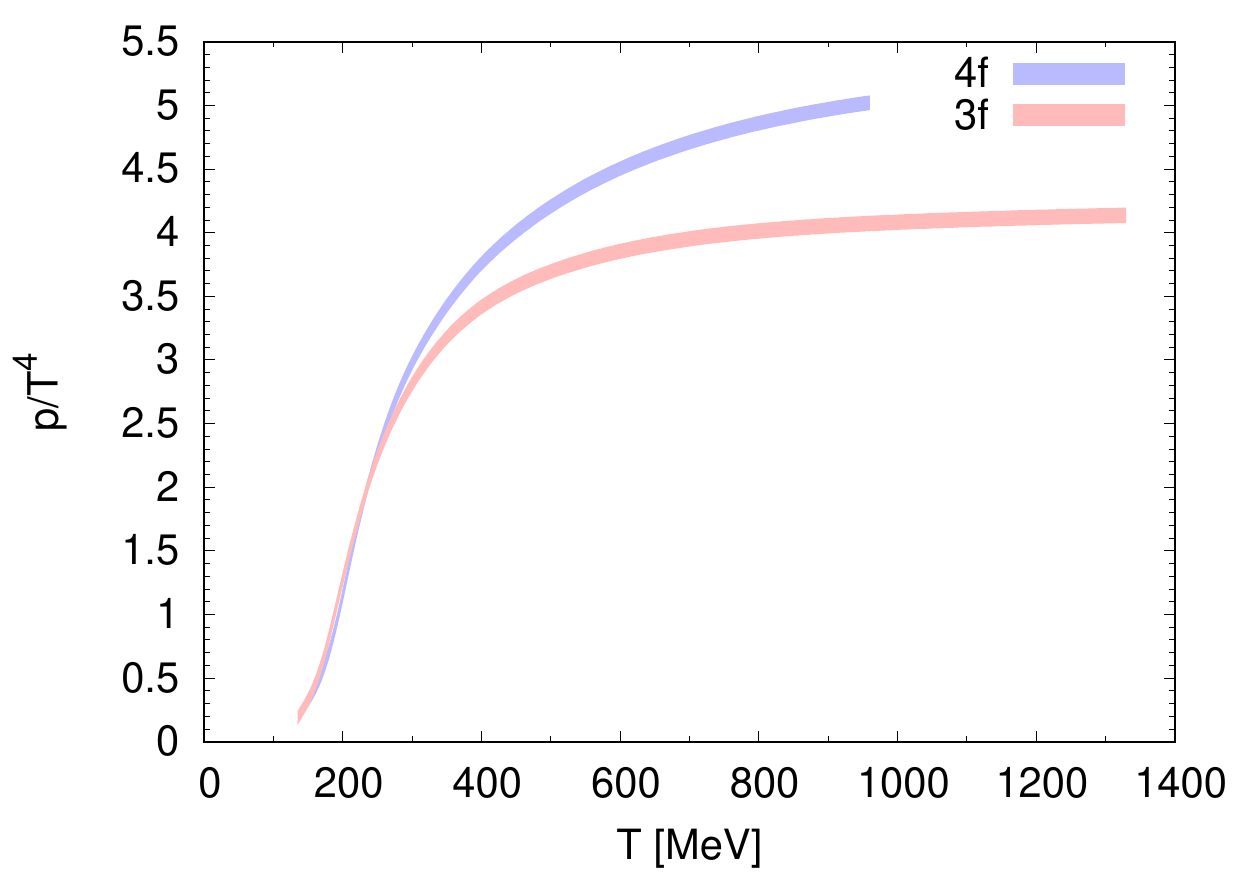}
\includegraphics[scale=0.5]{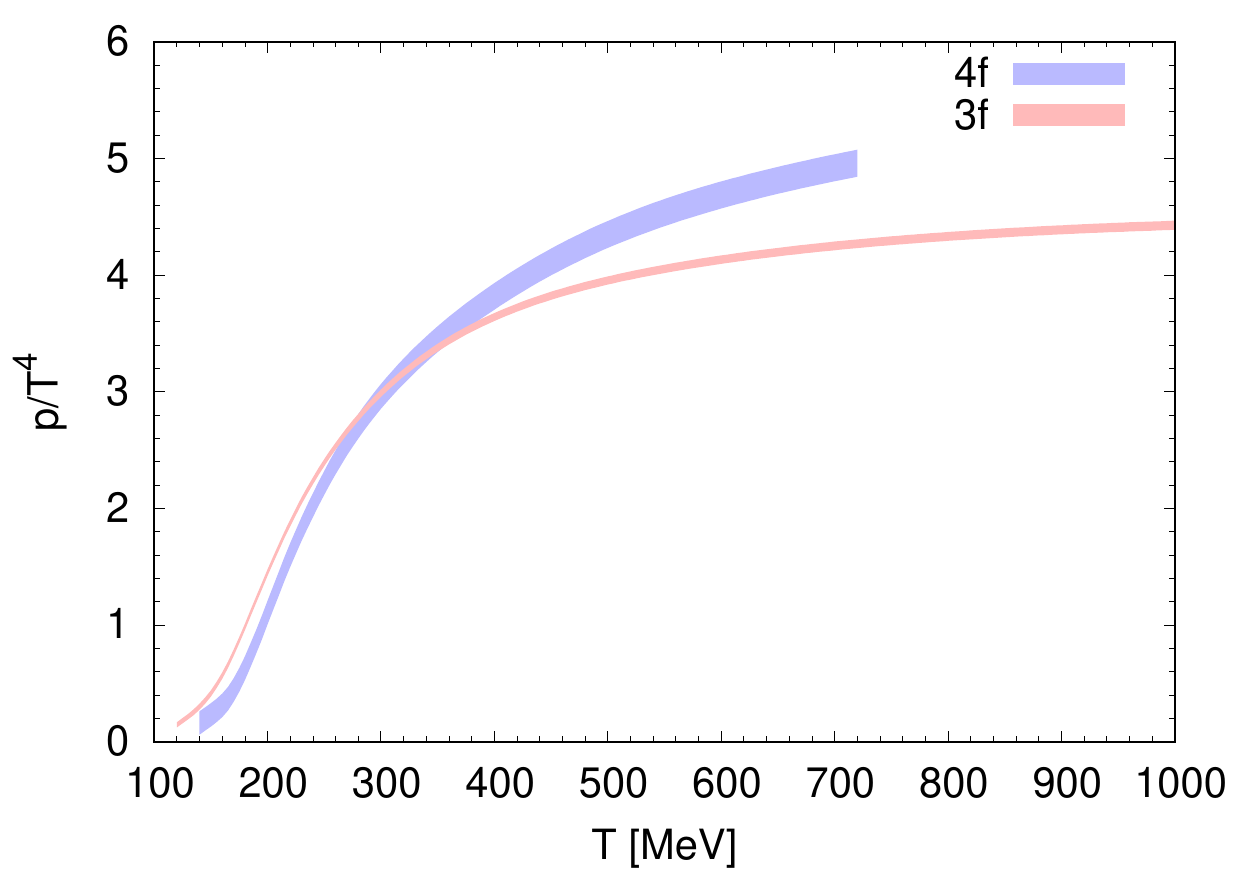}
\caption{
Pressure as function of temperature on $N_\tau=6$ (left) and $8$ (right) lattices for 
(2+1)- and (2+1+1)-flavor QCD. The errors are purely statistical. 
In the continuum limit, the (2+1)-flavor or (2+1+1)-flavor QCD results correspond to pion masses of $m_\pi \approx 160\,\mr{MeV}$ or $300\,\mr{MeV}$, respectively. 
\label{fig:3vs4fnt68}
}
\end{figure}

The gauge configurations are generated with the RHMC algorithm~\cite{Clark:2006fx}. 
At $T=0$ we save lattices every 5 or 6 and at $T>0$ every 10 molecular dynamics 
time units (TU).
The statistics for the $N_\tau=6$ or $8$ ensembles is reaching for most of them 50 thousand or
100 thousand TUs, respectively. 
In Fig. \ref{fig:e-3p} we show our results for the trace anomaly for different $N_{\tau}$.
As one can see from the figures we have accurate results for the trace anomaly
on $N_{\tau}=6$ and $N_{\tau}=8$ lattices. On the other hand there are large
fluctuations in the results obtained for $N_{\tau}=10$ and $12$.
Nonetheless, there is no apparent cutoff dependence of the trace anomaly for $N_{\tau}>6$.
Since we have accurate results for the trace anomaly for $N_{\tau}=6$ and $8$ 
we interpolate them with splines and then 
evaluate the pressure via the integral method as discussed above.
In Fig.~\ref{fig:3vs4fnt68} we compare the pressure in (2+1+1)-flavor QCD along 
the line of constant physics $m_\pi\approx 300\,\mr{MeV}$ with the pressure in 
(2+1)-flavor QCD along the line of constant physics 
$m_\pi\approx 160\,\mr{MeV}$~\cite{Bazavov:2017dsy}. 
Note that due to the difference in the pion mass for $N_{\tau}=8$ the (2+1+1)-flavor pressure is 
below the (2+1)-flavor pressure at low temperatures, $T \lesssim 300\,\mr{MeV}$, 
where the contribution of the charm quark is still negligible.
For $N_{\tau}=6$ we do not see significant differences since the cutoff effects
are more prominent than the quark mass effects.

\section{Conclusions}

We have extended the calculation of the equation of state in (2+1+1)-flavor 
QCD with  HISQ action and concluded the calculation on the coarse lattices.  
We have generated several new ($T=0$ and $T>0$) ensembles to achieve better 
coverage of the temperature range 130 - 1000~MeV and increased 
the statistics on most of the ensembles. 
We have reached lattice spacings down to $0.034\,\mr{fm}$, which corresponds to 
$967\,\mr{MeV}$ for $N_\tau=6$. Calculations on $N_{\tau}=6$ and $8$ lattices
shows that there is a significant contribution from charm quarks to the pressure
for $T>300$ MeV. However,
substantial increase in the statistics on the finer ensembles ($N_\tau=10,~12$) will be needed to accomplish a robust 
continuum extrapolation.

\section*{Acknowledgments}

The simulations have been carried out at NERSC and at the ICER of Michigan State University. 
This work is supported by the US Department of Energy, Office of Science, 
Office of Nuclear Physics: 
(i) Through the Contract No. DE-SC0012704; 
(ii) Through the Scientific Discovery through Advanced Computing (ScIDAC) 
award ``Computing the Properties of Matter with Leadership Computing Resources'';
(iii) Through the NSF award PHY-1812332.
J.H.W.’s research was also funded by Deutsche Forschungsgemeinschaft 
(DFG, German Research Foundation) -- 
Projektnummer 417533893/GRK2575 ``Rethinking Quantum Field Theory''. 

\bibliographystyle{JHEP}
\bibliography{ref.bib}

\providecommand{\href}[2]{#2}\begingroup\raggedright\begin{thebibliography}{10}

\bibitem{HotQCD:2018pds}
{\scshape HotQCD} collaboration, A.~Bazavov et~al., \emph{{Chiral crossover in
  QCD at zero and non-zero chemical potentials}},
  \href{https://doi.org/10.1016/j.physletb.2019.05.013}{\emph{Phys. Lett. B}
  {\bfseries 795} (2019) 15}
  [\href{https://arxiv.org/abs/1812.08235}{{\ttfamily 1812.08235}}].

\bibitem{Giusti:2016iqr}
L.~Giusti and M.~Pepe, \emph{{Equation of state of the SU(3)
  Yang\textendash{}Mills theory: A precise determination from a moving frame}},
  \href{https://doi.org/10.1016/j.physletb.2017.04.001}{\emph{Phys. Lett. B}
  {\bfseries 769} (2017) 385}
  [\href{https://arxiv.org/abs/1612.00265}{{\ttfamily 1612.00265}}].

\bibitem{Borsanyi:2013bia}
S.~Borsanyi, Z.~Fodor, C.~Hoelbling, S.~D. Katz, S.~Krieg and K.~K. Szabo,
  \emph{{Full result for the QCD equation of state with 2+1 flavors}},
  \href{https://doi.org/10.1016/j.physletb.2014.01.007}{\emph{Phys. Lett. B}
  {\bfseries 730} (2014) 99} [\href{https://arxiv.org/abs/1309.5258}{{\ttfamily
  1309.5258}}].

\bibitem{HotQCD:2014kol}
{\scshape HotQCD} collaboration, A.~Bazavov et~al., \emph{{Equation of state in
  ( 2+1 )-flavor QCD}},
  \href{https://doi.org/10.1103/PhysRevD.90.094503}{\emph{Phys. Rev. D}
  {\bfseries 90} (2014) 094503}
  [\href{https://arxiv.org/abs/1407.6387}{{\ttfamily 1407.6387}}].

\bibitem{Bazavov:2017dsy}
A.~Bazavov, P.~Petreczky and J.~H. Weber, \emph{{Equation of State in 2+1
  Flavor QCD at High Temperatures}},
  \href{https://doi.org/10.1103/PhysRevD.97.014510}{\emph{Phys. Rev. D}
  {\bfseries 97} (2018) 014510}
  [\href{https://arxiv.org/abs/1710.05024}{{\ttfamily 1710.05024}}].

\bibitem{Borsanyi:2016ksw}
S.~Borsanyi et~al., \emph{{Calculation of the axion mass based on
  high-temperature lattice quantum chromodynamics}},
  \href{https://doi.org/10.1038/nature20115}{\emph{Nature} {\bfseries 539}
  (2016) 69} [\href{https://arxiv.org/abs/1606.07494}{{\ttfamily 1606.07494}}].

\bibitem{Petreczky:2020tky}
P.~Petreczky and J.~H. Weber, \emph{{Strong coupling constant from moments of
  quarkonium correlators revisited}},
  \href{https://arxiv.org/abs/2012.06193}{{\ttfamily 2012.06193}}.

\bibitem{Komijani:2020kst}
J.~Komijani, P.~Petreczky and J.~H. Weber, \emph{{Strong coupling constant and
  quark masses from lattice QCD}},
  \href{https://doi.org/10.1016/j.ppnp.2020.103788}{\emph{Prog. Part. Nucl.
  Phys.} {\bfseries 113} (2020) 103788}
  [\href{https://arxiv.org/abs/2003.11703}{{\ttfamily 2003.11703}}].

\bibitem{Laine:2006cp}
M.~Laine and Y.~Schroder, \emph{{Quark mass thresholds in QCD thermodynamics}},
  \href{https://doi.org/10.1103/PhysRevD.73.085009}{\emph{Phys. Rev. D}
  {\bfseries 73} (2006) 085009}
  [\href{https://arxiv.org/abs/hep-ph/0603048}{{\ttfamily hep-ph/0603048}}].

\bibitem{MILC:2012aiy}
{\scshape MILC} collaboration, A.~Bazavov et~al., \emph{{Towards a QCD Equation
  of State with 2 + 1 + 1 Flavors using the HISQ Action}},
  \href{https://doi.org/10.22323/1.164.0071}{\emph{PoS} {\bfseries LATTICE2012}
  (2012) 071}.

\bibitem{MILC:2013ops}
{\scshape MILC} collaboration, A.~Bazavov et~al., \emph{{Update on the 2+1+1
  Flavor QCD Equation of State with HISQ}},
  \href{https://doi.org/10.22323/1.187.0154}{\emph{PoS} {\bfseries LATTICE2013}
  (2014) 154} [\href{https://arxiv.org/abs/1312.5011}{{\ttfamily 1312.5011}}].

\bibitem{Follana:2006rc}
{\scshape HPQCD, UKQCD} collaboration, E.~Follana, Q.~Mason, C.~Davies,
  K.~Hornbostel, G.~P. Lepage, J.~Shigemitsu et~al., \emph{{Highly improved
  staggered quarks on the lattice, with applications to charm physics}},
  \href{https://doi.org/10.1103/PhysRevD.75.054502}{\emph{Phys. Rev. D}
  {\bfseries 75} (2007) 054502}
  [\href{https://arxiv.org/abs/hep-lat/0610092}{{\ttfamily hep-lat/0610092}}].

\bibitem{Bazavov:2017lyh}
A.~Bazavov et~al., \emph{{$B$- and $D$-meson leptonic decay constants from
  four-flavor lattice QCD}},
  \href{https://doi.org/10.1103/PhysRevD.98.074512}{\emph{Phys. Rev. D}
  {\bfseries 98} (2018) 074512}
  [\href{https://arxiv.org/abs/1712.09262}{{\ttfamily 1712.09262}}].

\bibitem{MILC:2010hzw}
{\scshape MILC} collaboration, A.~Bazavov et~al., \emph{{Results for light
  pseudoscalar mesons}}, \href{https://doi.org/10.22323/1.105.0074}{\emph{PoS}
  {\bfseries LATTICE2010} (2010) 074}
  [\href{https://arxiv.org/abs/1012.0868}{{\ttfamily 1012.0868}}].

\bibitem{Boyd:1996bx}
G.~Boyd, J.~Engels, F.~Karsch, E.~Laermann, C.~Legeland, M.~Lutgemeier et~al.,
  \emph{{Thermodynamics of SU(3) lattice gauge theory}},
  \href{https://doi.org/10.1016/0550-3213(96)00170-8}{\emph{Nucl. Phys. B}
  {\bfseries 469} (1996) 419}
  [\href{https://arxiv.org/abs/hep-lat/9602007}{{\ttfamily hep-lat/9602007}}].

\bibitem{Allton:1996dn}
C.~R. Allton, \emph{{Lattice Monte Carlo data versus perturbation theory}},
  \href{https://doi.org/10.1016/S0920-5632(96)00804-3}{\emph{Nucl. Phys. B
  Proc. Suppl.} {\bfseries 53} (1997) 867}
  [\href{https://arxiv.org/abs/hep-lat/9610014}{{\ttfamily hep-lat/9610014}}].

\bibitem{Clark:2006fx}
M.~A. Clark and A.~D. Kennedy, \emph{{Accelerating dynamical fermion
  computations using the rational hybrid Monte Carlo (RHMC) algorithm with
  multiple pseudofermion fields}},
  \href{https://doi.org/10.1103/PhysRevLett.98.051601}{\emph{Phys. Rev. Lett.}
  {\bfseries 98} (2007) 051601}
  [\href{https://arxiv.org/abs/hep-lat/0608015}{{\ttfamily hep-lat/0608015}}].

\end{thebibliography}\endgroup

\end{document}